\documentclass[conference]{IEEEtran}

\usepackage{amsfonts}
\usepackage{amssymb}
\usepackage{algorithm,algorithmic}

\usepackage{graphicx}

\begin{document}

\title{Cache-aware static scheduling for hard real-time multicore
  systems based on communication affinities}

\author{\IEEEauthorblockN{Lilia Zaourar and Mathieu Jan}
\IEEEauthorblockA{CEA, LIST\\
Embedded Real Time Systems Laboratory\\
F-91191 Gif-sur-Yvette, France\\
Email: Firstname.Lastname@cea.fr}
\and
\IEEEauthorblockN{Maurice Pitel}
\IEEEauthorblockA{Schneider Electric Industries\\
37, quai Paul Louis Merlin\\
F-38050 Grenoble, France\\
Email: Maurice.Pitel@schneider-electric.com}}

\maketitle

\thispagestyle{empty}

\begin{abstract}
The growing need for continuous processing capabilities has led to the
development of multicore systems with a complex cache hierarchy. Such
multicore systems are generally designed for improving the performance
in average case, while hard real-time systems must consider worst-case
scenarios. An open challenge is therefore to efficiently schedule hard
real-time tasks on a multicore architecture. In this work,
we propose a mathematical formulation for computing a static
scheduling that minimize $L_{1}$ data cache misses between hard
real-time tasks on a multicore architecture using communication
affinities.
\end{abstract}

\section{Introduction}
\label{sec:intro}

\begin{figure}[!h]
\begin{center}
\includegraphics[scale=0.3]{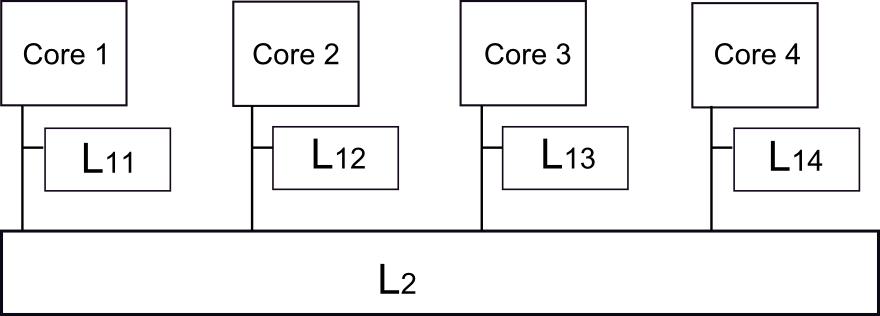}
\caption{Classical cache memory hierarchy of a multicore architecture.}
\label{cache-archi}
\end{center}
\end{figure}

Multicore processors have become the norm in many execution platforms
in various fields. Such architectures come with a cache memory
hierarchy made of several levels, shared or not between
cores. Figure~\ref{cache-archi} represents the typical cache hierarchy
that can be found in a multicore architecture with two levels of
cache, noted $L_{1x}$ and $L_{2}$ where $x$ is the number of the core
(in the figure $x$ ranges from 1 to 4). In such architectures, the
$L_{2}$ cache is larger but provides slower access time than the
$L_{1}$ cache.  Generally, the $L_2$ cache is unified and shared
among all cores, while at the $L_1$ level data and instruction
caches are separated and private to each core. However, in this work
we focus only on the data caches.
\vspace{-0.5cm}

Using such multicore architectures for developing hard real-time
systems is an important research area. Multicore are generally tuned
for optimizing performance for the average case, while hard real-time
systems must consider worst-case scenarios due to certification
constraints. A major problem lies in the management of cache for
mastering the impact of conflicts on the Worst-Case Execution Time
(WCET) of each task. Designing cache-aware scheduling policies is
becoming a popular research area. In this work, we show \emph{how a
  static scheduling that minimizes $L_{1}$ data cache misses between hard
  real-time tasks on a multicore architecture can be computed}. 



\section{Related work}
\label{sec:rw}

\cite{anderson2005real} focuses on the memory-to-$L_{2}$ traffic in the
cache hierarchy of soft real-time systems. They propose a two steps
method to discourage the co-scheduling of the tasks generating such
traffic. First, the tasks that may induce significant memory-to-$L_{2}$
traffic are gathered into groups. Then at runtime, they use a
scheduling policy that reduces concurrency within
groups. \cite{ECRTS08} also proposes several global multi-core
scheduling strategies for soft real-time systems to minimize the $L_{2}$
cache trashing. Co-scheduling of the tasks of a same group is used to
optimize the efficient use of the $L_{2}$ shared cache. Task promotion is
another example of a studied scheduling policy.
%
%

When considering hard real-time systems, to the best of our knowledge
we are only aware of~\cite{Guan2009}. Cache-partitioning techniques
are used to avoid interferences, at the $L_{2}$ cache level, between
the tasks that are running simultaneously. In addition to regular
temporal constraints used within a schedulability test, cache
constraints due to cache-partitionning are added and steer the
computation of the scheduling. They propose a linear programming
formulation to solve this problem and an approximation of this
formulation for larger task sets.

The closest work to ours is~\cite{lindsay2012lwfg}. While proposed
cache-aware scheduling strategies are evaluated using a soft-real time
kernel,
the results can also be used for hard real-time systems. They propose
a bin packing approach to evenly distribute the Working Set Size (WSS)
of the tasks on all cores in order to reduce conflicts.
The resolution algorithm is based on the next fit decreasing heuristic
applied on the tasks ordered by their decreasing WSS. Besides, they
rely on a notion of distance between caches of non-uniform memory
architectures to further optimize the solution. This is only used for
the tasks that share some common memory area and are gathered into
groups. However, it is unclear how the common memory area defines a
group as well as how groups are reduced when the heuristic fails to
allocate a group.

To summarize, most of the existing cache-aware scheduling proposal
have focused on the efficient use of the $L_{2}$ cache.

\section{Task model and notations}

Let $\Gamma = \left\lbrace \tau_{1},\tau_{2},...,
\tau_{n}\right\rbrace $ be a set of $n$ independent, synchronous,
preemptible and periodic tasks. $\Gamma$ is handled using the implicit
deadline periodic task model. Each task $\tau_{i} \in \Gamma$ has the
following temporal parameters $\tau_{i}=\left(P_{i}, C_{i}
\right)$. $P_{i}$ is the period of the task and $C_{i}$ is the WCET. A
job $i$ represents an instance of a task with $C_{i}$ its WCET. Let $H$
be the hyper-period of task set. It equals to the least common
multiple of all periods of tasks in $\Gamma$.

As in~\cite{lemerre08}, the hyper-period $H$ is divided in intervals,
an interval being delimited by two task releases. A job can be present
on several intervals, and we note $w_{j,k}$ the weight of job $j$ on
interval $k$. We denote by $I$ the set of intervals and $|I_{k}|$ the
duration of the $k^{th}$ interval, $|I_{k}|= t_{k+1} - t_{k}$. The
weight of each job is the amount of processor necessary to execute job
$i$ on interval $|I_{k}|$ only (it is not an execution time but a
fraction of it). $J_{\Gamma}$ is the job set of all jobs of $\Gamma $
scheduled during the hyper-period $H$.

Then, temporal schedulability constraints are expressed using a linear
program described in~\cite{megel2010minimizing} to compute the optimal
job weights on each interval for all tasks $T_{i} \in \Gamma$. First,
the sum of all job weights on an interval does not exceed the
processor maximum capacity:
\begin{equation}
\displaystyle \sum_{i \in J_{k}}w_{i,k} \leq M, \forall k
\end{equation} 
Then each job weight does not exceed each processor maximum capacity:
\begin{equation}
0 \leq w_{i,k} \leq 1 , \forall k, \forall i.
\end{equation} 
Finally, jobs must be completely executed: 
\begin{equation}
\displaystyle \sum_{k\in E{i}}w_{i,k} \times |I_{k}| = C_{i}, \forall i.
\end{equation} 

\section{Problem formulation}
\label{sec:pb}

The problem we address in this work is to reduce $L_1$ data cache
misses when scheduling hard real-time tasks on a multicore
architecture. To this end, we maximize the co-scheduling on a same
core of tasks that exchange data while still ensuring temporal
schedulability constraints. We assume a static knowledge of data
exchange between the tasks of an application. Therefore, we extend the
classical periodic task model with the WSS parameter for each task and
model this problem using a variant of the knapsack problem.  We also
assume that the system is schedulable and hence we only seek to
optimize the allocation of the tasks on the $L_1$ caches. In addition,
we suppose that the size of a $L_1$ cache enables to host several
tasks simultaneously, a valid hypothesis in the case studies we
consider. We leave as future work the management of cache conflicts,
using techniques such as in~\cite{WardHKA13}.

The multicore platform is made of $m$ cores and we note $C_{L_{1}}$
the capacity of each data cache $L_{1}$ (we assume the $L_{1}$ caches to
have an equal size). Finally, from the data of the application we can
calculate $WSS_{i}$ which is the WSS of job $J_{i}$. This is computed
by adding the size of each data section from the binary of an
application. We note $a_{ii^{'}}$ the affinity between the job $i$ and
the job $i^{'}$. The affinity between two tasks is defined as the
number of communication flows between them. Higher is the number of
communication flow, higher is the affinity between two
tasks. Communication flows between tasks are extracted using the
software architecture of the considered hard real-time application.

%

Then, integrating into the previously described schedulability
constraints require to introduce a decision variable representing the
allocation of the tasks on cores for each intervals. Let $x_{ijk}$ be
this decision variable that is equal to $1$ if the job $i$ is assigned
to cache $j$ during time interval $I_{k}, k\in \left\lbrace 1,...,
T\right\rbrace$ and $0$ otherwise. The sum of the WSS of the jobs
allocated to a cache should not exceed its capacity:
\begin{equation} 
\label{cap-cont}
 \displaystyle \sum_{i=1}^{n} WWS_{i} \times x_{ijk} \leq C_{L_{1}}, \forall j, \forall k. 
 \end{equation} 
In addition, each job must be assigned to a single cache:
\begin{equation}
\label{affect-cont}
\displaystyle\sum_{j=1}^{m} x_{ijk} \leq 1, \forall i, \forall k 
\end{equation} 
Besides, to link the temporal schedulability constraints with the
aforementioned cache constraints, the following relationship can be
define: if $\sum_{i=1}^{n} x_{ijk} =0$ then $w_{i,k} =0$ and if
$\sum_{i=1}^{n}x_{ijk} =1$ then $w_{i,k} > 0$. Finally, since our aim
is to maximise affinity $L_{1}$, we obtain the following objective
function:
\begin{equation}
\label{fct-obj}
  Max(Z)= \displaystyle \sum_{i=1}^{n}\sum_{j=1}^{m}\sum_{k=1}^{T}a_{ii^{'}} \times x_{ijk} \times x_{i^{'}jk}, \forall i, \forall j, \forall k.
 \end{equation} 

\section{Conclusion and future work}
\label{sec:conclusion}

In this work, we show how we can extend classical (temporal)
schedulability constraints to minimize $L_{1}$ data cache miss between
communicating hard real-time tasks on a multicore architecture. In
future work, we plan to generalise the formulation to address other
level of a cache memory hierarchy. As our formulation of the problem
uses a \textit{quadratic knapsack}, known to be
\textit{NP-hard}~\cite{kellerer2004knapsack}, another next step is
therefore the linearization of the objective function. Then, we plan
to implement it using the CPLEX solver to generate the static
scheduling of hard real-time tasks. Finally, we plan to test our
method on several hard real-time industrial applications.

\tiny


\end{document}